\documentstyle[11pt,saltwshop,twoside,epsf]{article}
\def\edcomment#1{\iffalse\marginpar{\raggedright\sl#1\/}\else\relax\fi}

\reversemarginpar

\begin{document}
\title{Galaxy Kinematics with SALT}

\author{M. A. Bershady$^1$, M. A. W. Verheijen$^2$,
D. R. Andersen$^3$, R. A. Swaters$^4$, and K. B. Westfall$^1$}

\affil{$^1$Department of Astronomy, University of Wisconsin--Madison,
475 N. Charter St, Madison, WI 57306, USA \\ $^2$Astrophysikalisches
Institut Potsdam, An der Sternwarte 16, 14482, Potsdam, Germany \\
$^3$Max Planck Institut f\"{u}r Astronomie, K\"{o}nigstuhl 17, 69117
Heidelberg, Germany \\ $^4$Department of Astronomy, University of
Maryland, College Park, MD 20742, USA}

\begin{abstract}

The combination of dynamical and photometric properties of galaxies
offers a largely un-tapped source of information on how galaxies
assembled and where stars formed. Bi-dimensional kinematic
measurements have been the stumbling block. The light-gathering power
of SALT coupled with the high-throughput performance of the Prime
Focus Imaging Spectrograph (PFIS) yield a superb facility for
measuring velocity-ellipsoids of stars and gas in galaxies out to
gigaparsec distances. From these data dynamical asymmetries arising
from lopsided or elliptical halos may be probed; disk-mass and
mass-decompositions may be uniquely determined; mechanisms for disk
heating constrained; and a zeropoint for the mass-to-light ratios of
stellar populations set. A number of groups within the SALT consortium
are interested in making these measurements using a variety of
different, but complementary approaches. The scientific potential from
their synthesis is very promising. We describe some unusual
observational modes in which PFIS may be used to probe the shape of
dark-matter halos and the content of galaxy disks.

\end{abstract}

\section{Introduction}

Extant information about galaxies comes primarily from broad-band
optical images. These deliver a wealth of information on when and
where stars have formed. Missing is an accurate understanding of the
distribution of mass and the details of its assembly -- knowledge
which requires dynamical arguments substantiated with kinematic
measurements. Hence kinematic information is essential for a basic
understanding of galaxy formation and evolution. The last decade has
seen the growth of bi-dimensional optical spectroscopy -- once the
purview of radio (HI) observations. SALT can continue this
development. With SALT's Prime Focus Imaging Spectrograph we will be
able to measure velocity-ellipsoid maps for a variety of dynamical
tracers which include (collisionless) stellar populations and (`sticky')
ionized gas. These measurements will be accessible via Fabry-Perot and
dispersed spectroscopy over a very large (8$^\prime$) field.

\begin{figure}
\plotfiddle{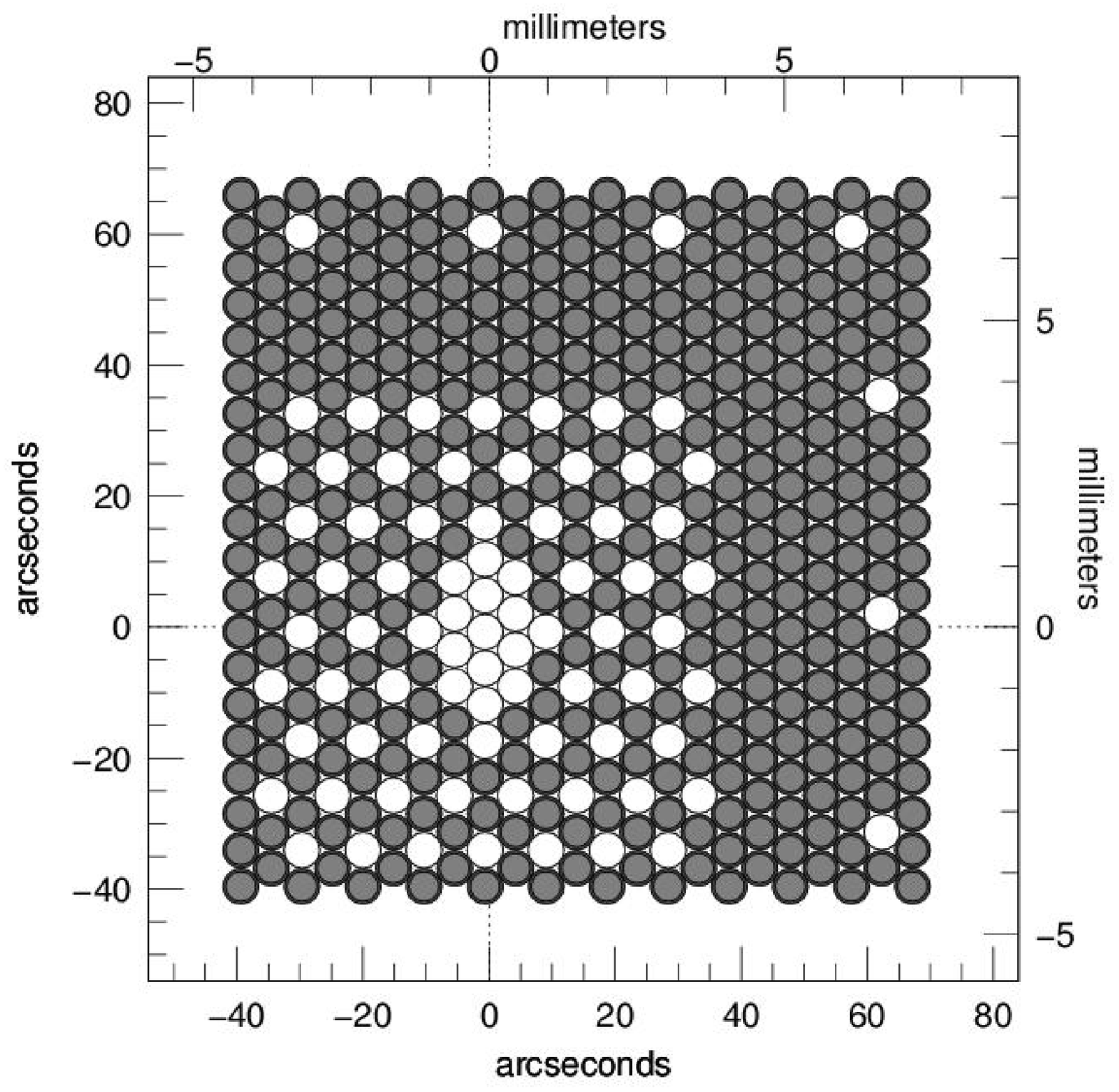}{2.0in}{0}{34}{34}{-200}{-18.5}
\plotfiddle{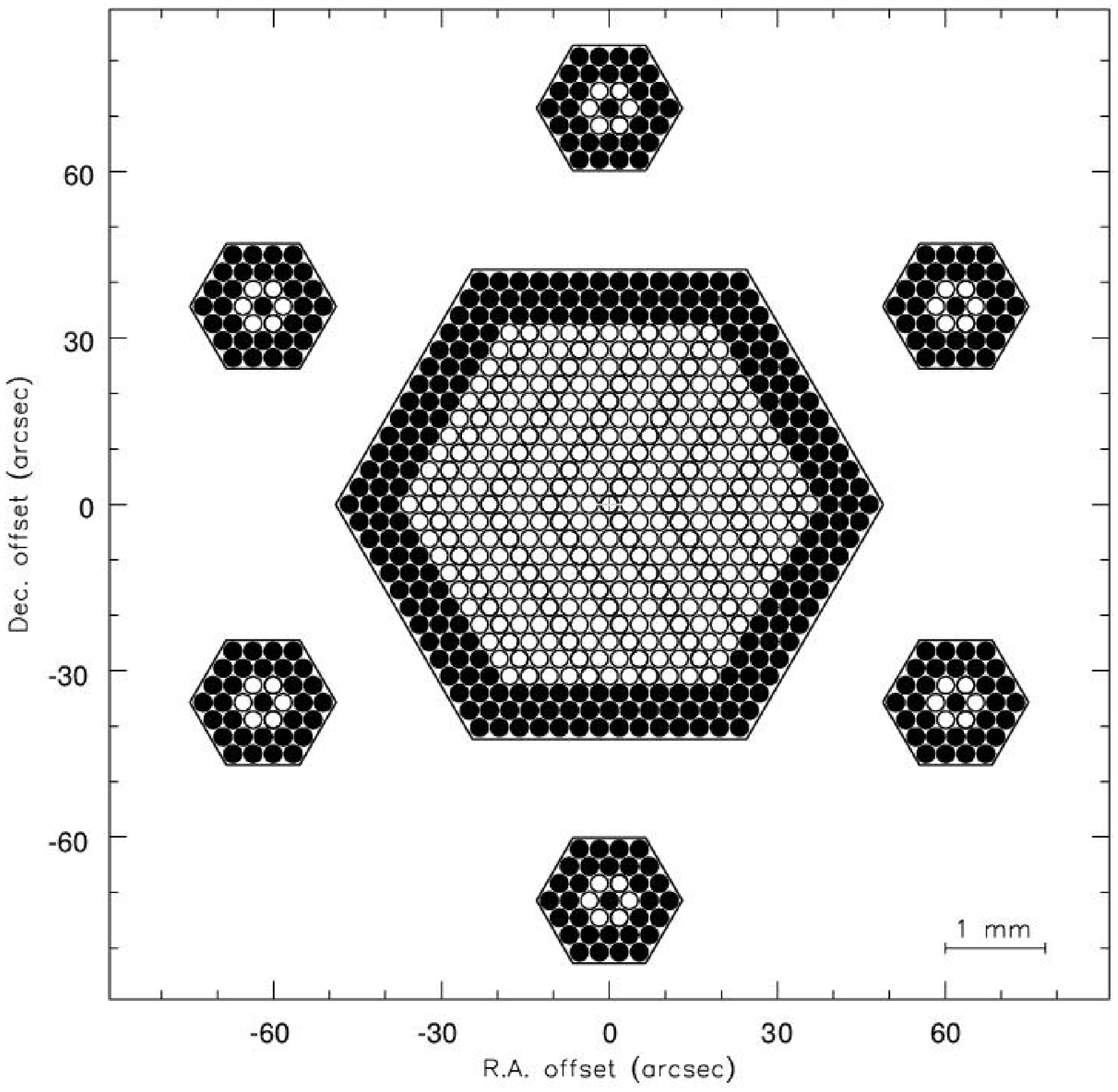}{0in}{0}{35.7}{35.7}{-10}{-16.5}
\caption{{\it SparsePak} ({\bf left}, Bershady et al. 2003, 2004) and
{\it PPAK} ({\bf right}, Verheijen et al. 2004) IFUs, which
respectively feed spectrographs on the WIYN and Calar Alto 3.5m
telescopes, both capable of $\sim$11 km/s ($\sigma$) resolution.
Units are presented at the same subtended angular scale at the
telescope focal-plane; physical dimensions scale with respective
telescope+fore-optic $f$-ratios. Light fibers transmit to the
spectrograph; dark fibers serve as mechanical buffers.}
\vskip -0.1in
\end{figure}

As promising for SALT is the critical mass of consortium scientists
with overlapping interests in galaxy dynamics (see, for example,
contributions by Balona, Cecil, Cote, Crawford, Ferrarese, Sellwood,
Sparke, Wilcots, Williams, and Ziegler in this volume). A few specific
scientific prospects of interest to members of the consortium include
(but are not limited to!) the frequency and amplitude of lopsided
disks and triaxial halos as probed by dynamical and photometric
asymmetries; mergers and perturbations inferred studies of kinematic
irregularities; disk heating mechanisms judged from velocity
dispersion ellipsoids ($\sigma_z/\sigma_R$, Gerssen et
al. 2000); feed-back processes in disks and the ISM; bar dynamics and
their pattern-speeds; the Tully-Fisher relation and its evolution; and
direct dynamical mass decompositions determined by combining rotation
curves with stellar velocity dispersions.

Here we focus on how well we can directly separate disk from halo mass
in spirals to yield dark-halo profiles, the zeropoint for stellar M/L
ratio, and plausible constraints on the faint-end of the stellar
IMF. These are of fundamental interest, respectively for testing both
hierarchical cold-dark-matter structure-formation scenarios and
star-formation models. This particular science case allows us to
illustrate a novel application of PFIS to bi-dimensional spectroscopy
with wide application. The concept combines narrow-band filters with
dispersed spectroscopy to form what we call {\it m}assively {\it
m}ulti-plexed {\it s}pectroscopy, or MMS. The approach complements
other methods using fluid-dynamical modeling and kinematic
measurements of barred potentials (Weiner et al. 2001; Gerssen et al.
2003; Debattista \& Williams 2004).

\section{Mass Decomposition}

\begin{figure}
\plotfiddle{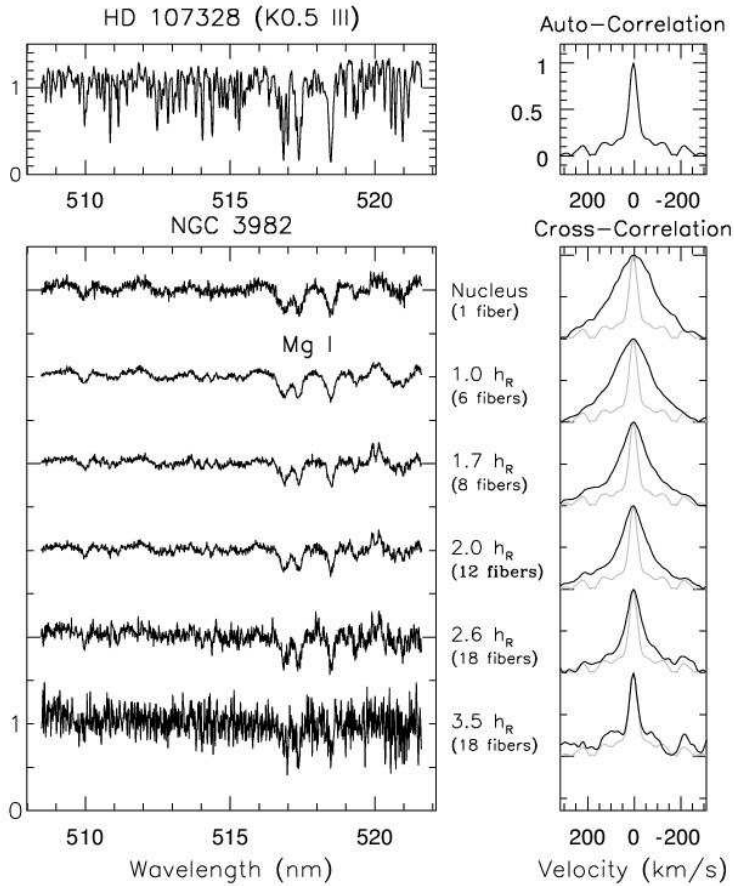}{3.85in}{0}{52.5}{52.5}{-150}{-45}
\vskip 0.3in
\caption{SparsePak spectra of a K0.5~III template-star and {\it
azimuthally-averaged} fiber rings for NGC 3982 at 6 different radii,
labeled in units of disk radial scale-length h$_R$.  Auto- and
cross-correlations are shown (right) for $\sim$10nm of spectra in the
MgI region.}
\end{figure}

The traditional technique of decomposing galaxy rotation curves into
disk and halo mass contributions is inherently degenerate (see
discussion by Sellwood, this volume). One way to break the disk-halo
degeneracy is to measure directly the disk potential via the vertical
motions and scale-height of the stars. To within factors of order
unity: $\sigma_z = \sqrt{\pi G \ z_0 \ \mu \ M/L}$, where $\sigma_z$
is the vertical component of the disk velocity dispersion, $z_0$ is
the disk scale height, $\mu$ the surface light-intensity, and M/L the
mass-to-light ratio (hence $\mu \ M/L$ is the disk surface
mas-density, $\Sigma$). Recent studies of edge-on galaxies (Kregel et
al. 2002) permit vertical scale-heights to be inferred with reasonable
accuracy for face-on systems given the observed correlations between
the radial scale-length, $h_R$, rotation speed, type, and $z_0$.
Face-on disks are ideal for measuring $\sigma_z$, typically the
smallest component of the velocity ellipsoid, but low inclination
makes measurement of inclination and rotation speed difficult.
Inclined disks require careful correction and decomposition of the
observed, line-of-sight velocity-dispersion to extract
$\sigma_z$. Past attempts to realize this kinematic approach (e.g.,
Bottema 1997) have not overcome these problems, and have been limited
by long-slit spectroscopy that does not reach well beyond a single
disk scale length.

\begin{figure}
\plotfiddle{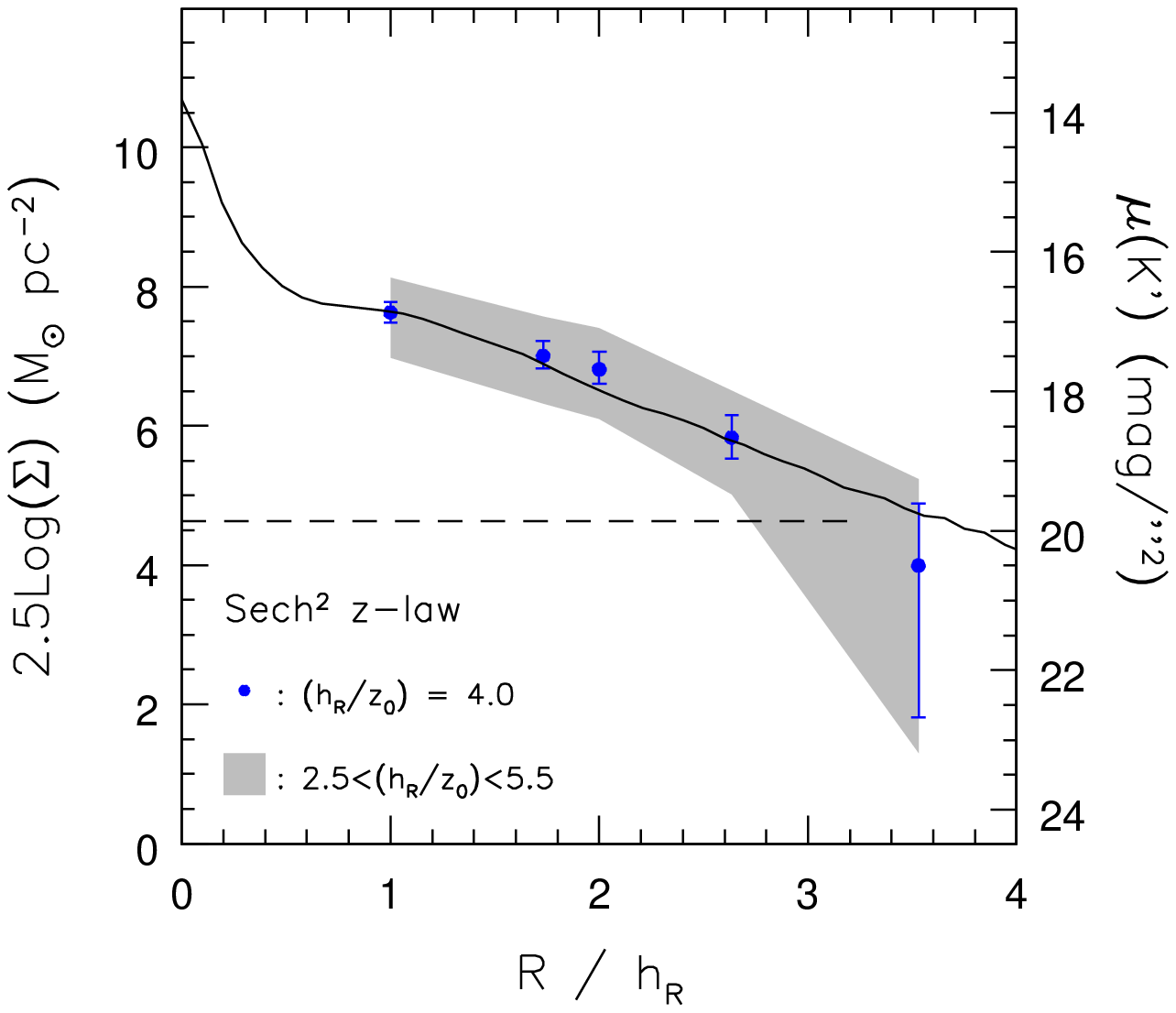}{1.750in}{0}{56}{56}{-269}{-185}
\plotfiddle{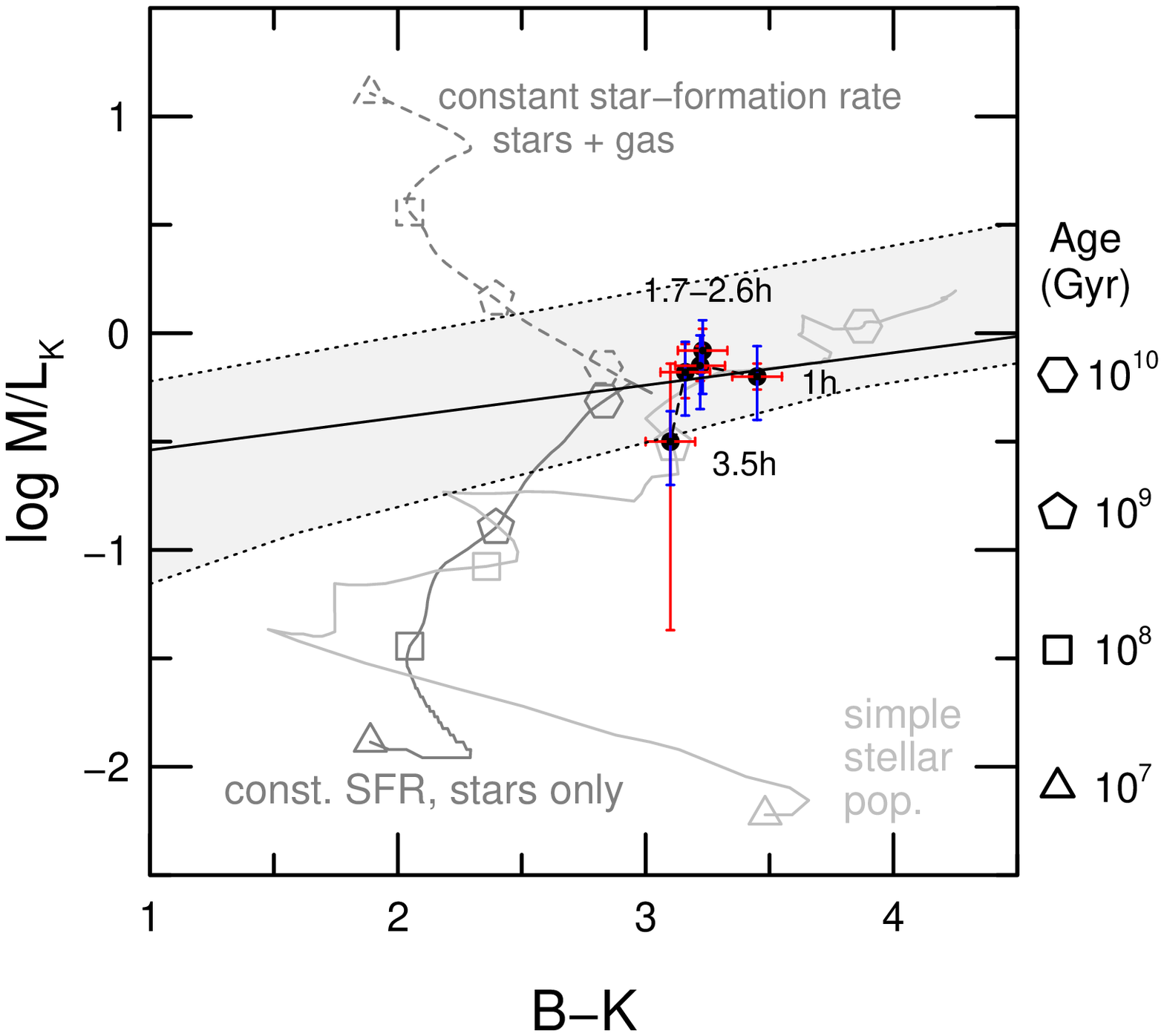}{0in}{0}{42}{42}{0}{-65}
\caption{Direct kinematic surface-mass density and M/L measurements
from integral-field spectroscopy at 515nm and optical-NIR photometry
of NGC 3982: Surface mass-density ($\Sigma$) and surface-brightness
($\mu$) vs radius ({\bf left}); $K$-band M/L vs rest-frame $B-K$ color
({\bf right}). Dashed line (left) represents a reference $\Sigma$ for the MW in
the solar neighborhood (Kuijken \& Gilmore 1991). Model
curves (right) are adopted from Bruzual \& Charlot (1993) and Bell \&
De Jong (2000). Error limits show systematic and random components.}
\end{figure}

The observational approach can be rectified by application of
integral-field observations of nearly face-on disks at medium spectral
resolutions ($7000 < \lambda / \Delta\lambda < 15000$) and
multi-band optical and NIR imaging. Two suitable integral-field units
(IFUs) are shown in Figure 1. Galaxy sizes can be carefully matched to
the IFUs. Observations with these units yield kinematic maps from
which accurate inclinations can be derived to $i \sim 15^\circ$
(Andersen \& Bershady 2003). Most importantly, the area sampled with
IFUs increases with radius. For nearly face-on disks, one may average
the signal in rings, which greatly enhances the limiting depth and
radius of the observations.

Using the two IFUs shown in Figure 1 we have undertaken a survey of
nearly face-on, nearby disk galaxies to determine their disk masses.
The survey has yielded H$\alpha$ kinematic maps for $\sim$100 normal,
late-type galaxies. We are now gathering absorption-line spectroscopic
observations (for stellar kinematics) for 40 of these galaxies,
selected to have regular velocity fields (see Verheijen et al. 2004,
Figure 2).

Results from a pilot target, NGC 3982, yield $\sigma_z$ out to 3.5
disk scale-lengths in this albeit high-surface-brightness disk in both
the MgI (515nm; Figure 2) and CaII triplet (860nm)
regions. Cross-correlations yield similar results and show that the
trend of disk surface-density well traces near-infrared
surface-brightness -- {\it mass traces light!} -- as shown in Figure
3. Contrary to our earlier, preliminary reports, a renewed analysis of
the data indicates a K-band M/L of the disk which is consistent with a
maximum-disk situation in the sense of van Albada \& Sancisi (1986).
The colors and dynamical M/L {\it measurements} in Figure 3 agree well
with stellar population synthesis models of, e.g., Bell \& De Jong
(2001), yet this source lies over 2 mag {\it below} TF. Here is a
small, blue high-surface-brightness, actively star-forming disk, which
is rapidly rotating and dominated by normal, baryonic matter out to
3-4 scale-lengths of the light distribution.  At face value, this
calibration of Bell \& De Jong's (2001) models implies, as they note,
a Salpeter-like IMF truncated below 0.35 M$_\odot$. It is critical to
test the generality of this result in a variety of ``normal'' systems
over a range in type, color and surface-brightness. Our on-going
survey will begin to address this issue, but the
low-surface-brightness regime will be difficult to probe at the
required spectral resolution using 4m-class telescopes.

\begin{figure}
\plotfiddle{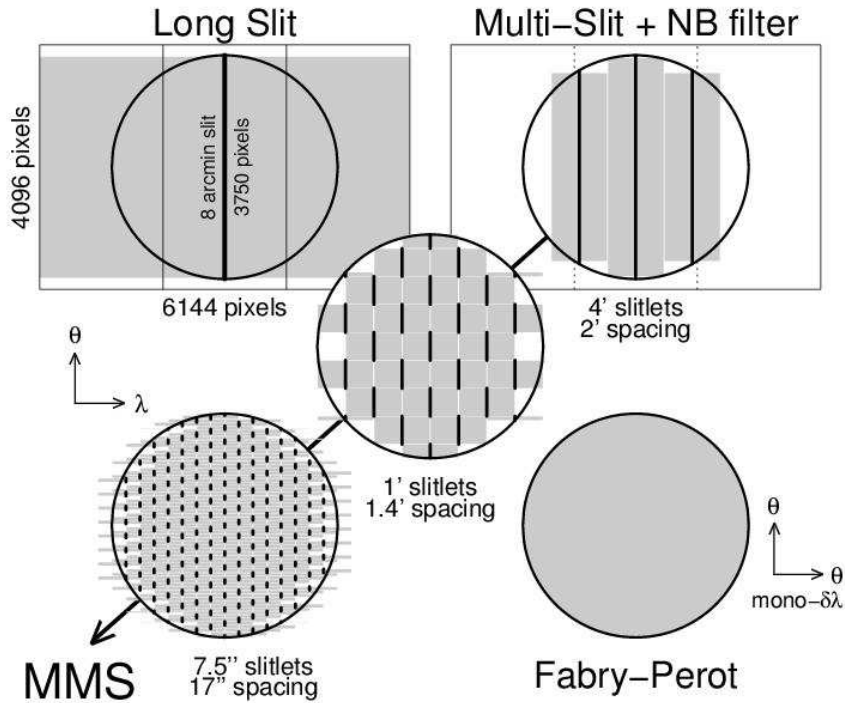}{2.0in}{0}{62}{62}{-180}{-124}
\vskip 1.3in
\caption{Examples of Massively Multiplexed Spectroscopy (MMS) compared
to Long-slit and Fabry-Perot spectroscopy with PFIS. Each present a
different slice of the spatial ($\theta$) $\times$ spatial ($\theta$)
$\times$ wavelength ($\lambda$) data-cube. ``NB'' is a narrow-band
filter, which in the case of PFIS has a bandwidth of
$\lambda/\Delta\lambda = 50$. Grey regions denote used portions of the
CCD. Long-slit spectroscopy makes best use of the CCD array, while MMS
and Fabry-Perot modes use only $\sim$50\% compared to the Long-slit
case.}
\vskip -0.05in
\end{figure}

\section{Prospects with the Prime Focus Imaging Spectrograph}

The power of PFIS lies in its unique combination of 3 Fabry-Perot
etalons spanning a factor of 50 in resolution, 6 gratings covering
from the UV-atmospheric cutoff to 850 nm at resolutions up to
$\lambda/\Delta\lambda \sim13000$, and a suite of narrow-band filters
filling an octave at optical wavelengths. All of these options are
configurable with imaging, longslit, or multi-slit mode. Most of the
gratings are high-transmissivity volume-phase holographic (VPH)
elements, tunable with an articulating camera.

For most galaxy kinematic studies, long-slit measurements are
inefficient: Two-dimensional sampling is needed to gain velocity field
information and to gather more light at large radii. Fabry-Perot
imaging, in contrast, is excellent for gas and stellar velocity fields
based on single, strong lines. However, for velocity dispersion work
in narrow-lined (low-mass) systems, ideally many (weak) lines are
probed, which requires scanning a significant band-pass
($\lambda/\Delta\lambda < 100$). Hence an intermediate trade between
the spectral and spatial sampling of longslit and Fabry-Perot
spectroscopy is needed. Integral- or formatted-field spectroscopy,
with limited spatial coverage is one alternative, but at first-blush
PFIS appears to have no such capability.

Other possibilities include using different tracers, such has
planetary nebulae (e.g., Ciardullo et al. 2003; Merrett et al. 2003).
Traditionally this requires narrow-band imaging and multi-object
spectroscopic follow-up. PFIS is well suited for both. While this is
efficient, newer techniques, such as counter-dispersed imaging (e.g.,
Douglas et al. 2002) also can be accomplished with PFIS. Another
approach for multi-object emission-line kinematics would be to use
slitless, dispersed imaging, i.e., slitless grism spectroscopy, at
high spectral resolution over a limited ``on'' and ``off'' band-pass
using a combination of gratings and narrow-band filters. The on and
off bands replace the function of counter-dispersion. The technique
should be considered by prospective PFIS users.  We describe, instead,
an alternative method using gratings and narrow-band filters with
slits for IFU-like spectroscopy -- suitable for emission and
absorption line studies of extended sources.

\subsection{MMS: Massively Multiplexed Spectroscopy}

The wavelength-spatial multiplex trade for galaxy kinematic studies
can be accomplished with PFIS by combining the suite of
$\lambda/\Delta\lambda \sim 50$ narrow-band filters with the tunable
VPH gratings.  The filters serve to limit the range of the dispersed
spectra on the detector, thereby allowing for an increase in the
spatial multiplex: Multiple, parallel slits or slit-lets can be placed
on a mask and produce non-overlapping spectra covering roughly, e.g.,
10nm centered at 515nm -- just what is shown in Figure 2 from
SparsePak. The concept for mapping extended sources is illustrated in
Figure 4; multi-object applications are possible (Crawford, this
volume).\footnote{The Fabry-Perot image is not truly monochromatic
(there is a radial field-dependence to the band-pass), and the MMS
slits are not uniformly spaced due to field-dependence of the grating
incidence angle.}

The minimum spatial separation in the dispersion direction is $\sim$1
arcmin at the highest spectral resolutions (grating angle
$\alpha=50^\circ$), and decreases linearly with resolution ($\propto
tan \ \alpha$, i.e., the Littrow condition for the VPH gratings). At
the highest spectral resolutions MMS yields a spatial multiplex
increase of roughly a factor of 4. {\bf For lower-dispersion setting,
the spatial multiplex can be significantly increased.} At any
resolution, the slits can be staggered to yield a variety of
two-dimension sampling patterns (Figure 4) that resemble e.g.,
SparsePak. Such data are suitable for constructing velocity fields and
dispersion maps.  Hence MMS should provide a competitive approach to
the stellar dynamical studies illustrated in the previous section.

It is useful to examine where PFIS lies in comparison to other
bi-dimensional spectrographs in terms of ``information gathering
power,'' which we loosely parametrize with the two quantities ``total
grasp'' and ``spectral power''. These parameters are defined in Figure
5, where it is seen that PFIS has very large grasp while spanning a
wide range in spectral power. The addition of MMS extends the range in
grasp and spectral power, bridging between Fabry-Perot and traditional
slit-spectroscopy modes. The light-gathering power of PFIS and SALT is
over an order of magnitude larger than the 4m-class IFU instruments we
have discussed. As such, PFIS can extend the study of galaxy
kinematics significantly into the low surface-brightness regime.

\section{Summary and Discussion}

\begin{figure}
\plotfiddle{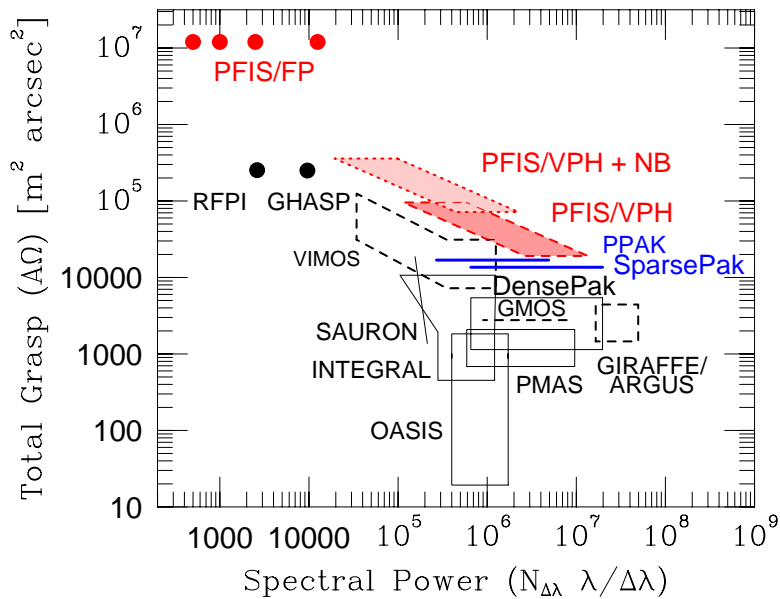}{2.75in}{-90}{75}{75}{-210}{265}
\caption{Grasp versus spectral power for two-dimensional spectroscopic
systems on 4m and 8m-class telescopes (solid and dashed lines,
respectively, with Fabry-Perot instruments shown as filled circles).
Total grasp is the product of area $\times$ solid-angle
(A$\Omega$). Spectral power is the product of the spectral resolution,
$\lambda/\Delta\lambda$, times the number of spectral resolution
elements, N$_{\Delta\lambda}$.  Variations in the shapes of covered
parameter space depends on how a given instrument achieves a range of
spectral resolution and spatial sampling, i.e., through changes in
gratings, slit-widths, or both.  Note the unique location of PFIS in
these diagrams is significantly extended with the use of MMS.}
\vskip -0.05in
\end{figure}

Galaxy kinematic measurements with SALT's PFIS can be used to answer
many outstanding, and fundamental questions about the structure and
formation of galaxies. Highlighted here is one example of unique,
dynamical mass decompositions of spiral galaxies. A number of
observational modes are available with PFIS. We have described one
powerful, new mode, referred to as ``massively multiplexed
spectroscopy,'' (MMS), which is enabled by the unique combination of
VPH gratings and narrow-band filters in PFIS. With MMS and Fabry-Perot
imaging, PFIS is a highly competitive survey instrument for
two-dimensional studies of gas {\it and} stars in nearby
galaxies. There are several groups within the SALT consortium
interested in applying these to a similar range of dynamical
problems. This holds the promise of fruitful collaboration,
cross-checks, and innovation.

What about higher redshift studies (see Ziegler and Crawford, this
volume)? PFIS can sample a 0.6 arcsec slitwidth, but the sub-arcsec
image quality enjoyed regularly by VLT and Gemini is needed unless we
remain content with spatially-unresolved line-widths.  Should it prove
feasible to phase the SALT primary segements, we may consider pushing
aggressively for significantly improved image quality. This will have
a major impact on both resolved, extragalactic as well as
high-resolution stellar spectroscopic observations. But SALT is merely
a 9m telescope; for extended sources observed at significant spectral
resolution, the area--solid-angle product (A$\Omega$) is what
counts. As other facilities rush for the highest possible angular
resolution that, with today's detectors, may be better suited for
tomorrow's 30m-class telescopes, we should keep good aim on the type
of ground-breaking science enabled by our large field of view and the
exquisite, multi-faceted spectroscopic capabilities of PFIS.

\acknowledgments This work is funded by NSF AST-0307417, NASA LTSA
NAG5-6032 and STScI/GO-9126.

\end{document}